\documentclass{ws-procs975x65}

\begin{document}

\title{QUARK SPIN IN THE PROTON}

\author{STEVEN D. BASS}

\address{Institute for Theoretical Physics, University of Innsbruck, \\
Technikerstrasse 25, A6020 Innsbruck, Austria \\
Steven.Bass@uibk.ac.at}

\begin{abstract}
The proton spin puzzle has challenged our understanding of QCD for the 
last 20 years.
We survey new developments in theory and experiment.
The proton spin puzzle 
seems to be telling us about the interplay of valence quarks with chiral 
dynamics and
the complex vacuum structure of QCD.
\end{abstract}


\bodymatter

\section{Introduction}

Protons behave like spinning tops.
Unlike classical tops, however, the spin of these particles is an
intrinsic quantum mechanical phenomenon.
This spin is responsible for many fundamental properties of matter,
including the proton's magnetic moment,
the different phases of matter in low-temperature physics,
the properties of neutron stars,
and the stability of the known universe.
How is the proton's spin built up from 
the spin and orbital angular momentum of the quarks and gluons inside ?

Polarized deep inelastic scattering experiments have revealed a small
value for the nucleon's flavour-singlet axial-charge
$g_A^{(0)}|_{\rm pDIS} \sim 0.3$
suggesting that
the quarks' intrinsic spin contributes little of the proton's spin.
The challenge to understand the spin structure of the
proton~\cite{bassrmp,bassbook}
has inspired a vast programme of theoretical activity and new experiments.
Why is the quark spin content $g_A^{(0)}|_{\rm pDIS}$ so small ?

We start by recalling the $g_1$ spin sum-rules, 
which are
derived starting from the dispersion relation for polarized
photon-nucleon scattering and, for deep inelastic scattering,
the light-cone operator product expansion.
One finds that the first moment of 
the $g_1$ spin structure function
is related
to the scale-invariant axial-charges of the target nucleon by
\begin{eqnarray}
\int_0^1 dx \ g_1^p (x,Q^2)
&=&
\Biggl( \frac{1}{12} g_A^{(3)} + \frac{1}{36} g_A^{(8)} \Biggr)
\Bigl\{1 + \sum_{\ell\geq 1} c_{{\rm NS} \ell\,}
\alpha_s^{\ell}(Q)\Bigr\}
\nonumber \\
& &
+ \frac{1}{9} g_A^{(0)}|_{\rm inv}
\Bigl\{1 + \sum_{\ell\geq 1} c_{{\rm S} \ell\,}
\alpha_s^{\ell}(Q)\Bigr\}  +  {\cal O}(\frac{1}{Q^2})
 + \ \beta_{\infty}
.
\nonumber \\
\label{eqc50}
\end{eqnarray}
Here $g_A^{(3)}$, $g_A^{(8)}$ and $g_A^{(0)}|_{\rm inv}$ are the
isovector, SU(3) octet and scale-invariant  flavour-singlet 
axial-charges respectively. 
The flavour non-singlet $c_{{\rm NS} \ell}$
and singlet $c_{{\rm S} \ell}$ Wilson coefficients are calculable in
$\ell$-loop perturbative QCD \cite{Larin:1997}.
The term $\beta_{\infty}$
represents a possible leading-twist subtraction constant
from the circle at infinity when one closes the contour in the
complex plane in the dispersion relation \cite{bassrmp}.
If finite, the subtraction constant affects just the first moment.
The first
moment of $g_1$ plus the subtraction constant, if finite, is equal
to the axial-charge contribution.

In terms of the flavour dependent axial-charges
\begin{equation}
2M s_{\mu} \Delta q =
\langle p,s |
{\overline q} \gamma_{\mu} \gamma_5 q
| p,s \rangle
\label{eqc55}
\end{equation}
the isovector, octet and singlet axial-charges are:
\begin{eqnarray}
g_A^{(3)} &=& \Delta u - \Delta d
\nonumber \\
g_A^{(8)} &=& \Delta u + \Delta d - 2 \Delta s
\nonumber \\
g_A^{(0)}|_{\rm inv}/E(\alpha_s)
\equiv
g_A^{(0)}
&=& \Delta u + \Delta d + \Delta s
.
\label{eqc56}
\end{eqnarray}
Here
$
E(\alpha_s) = \exp \int^{\alpha_s}_0 \! d{\tilde \alpha_s}\,
\gamma({\tilde \alpha_s})/\beta({\tilde \alpha_s})
$
is a renormalization group factor which
corrects for the (two loop) non-zero anomalous dimension $\gamma(\alpha_s)$
of the singlet axial-vector current~\cite{kodaira},
$
J_{\mu5} =
\bar{u}\gamma_\mu\gamma_5u
                  + \bar{d}\gamma_\mu\gamma_5d
                  + \bar{s}\gamma_\mu\gamma_5s 
$ ,
and which goes to one in the limit
$Q^2 \rightarrow \infty$;
$\beta (\alpha_s)$ is the QCD beta function.
The singlet axial-charge, $g_A^{(0)}|_{\rm inv}$,
is independent of the renormalization scale $\mu$
and corresponds
to
$g_A^{(0)}(Q^2)$ evaluated in the limit $Q^2 \rightarrow \infty$.
The axial-charges 
$g_A^{(3)}$ and  $g_A^{(8)}$ are renormalization group invariants.

If one assumes no twist-two subtraction constant
($\beta_{\infty} = O(1/Q^2)$)
the axial-charge contributions saturate the first moment
at leading twist.
The isovector axial-charge is measured independently in neutron
$\beta$-decays
($g_A^{(3)} = 1.270 \pm 0.003$ \cite{PDG:2004})
and the octet axial-charge is commonly taken
to be the value extracted
from hyperon $\beta$-decays assuming a
2-parameter SU(3) fit
($g_A^{(8)} = 0.58 \pm 0.03$ \cite{fec}). 
Using the sum-rule for the first moment of $g_1$, given in Eq.~(1),
polarized deep inelastic scattering experiments have been
interpreted in terms of
a small value for the flavour-singlet axial-charge.
If we take $g_A^{(8)} = 0.58 \pm 0.03$,
then
inclusive $g_1$ data with $Q^2 > 1$ GeV$^2$
give~\cite{compassnlo}
\begin{equation}
g_A^{(0)}|_{\rm pDIS, Q^2 \rightarrow \infty}
=
0.33 \pm 0.03 ({\rm stat.}) \pm 0.05 ({\rm syst.})
\end{equation}
-- considerably smaller than the value of $g_A^{(8)}$ quoted above.

In the naive parton model $g_A^{(0)}|_{\rm pDIS}$ is interpreted
as the fraction of the proton's spin which is carried by the intrinsic
spin of its quark and antiquark constituents.
When combined with
$g_A^{(8)} = 0.58 \pm 0.03$
the value of $g_A^{(0)}|_{\rm pDIS}$ in Eq.(4) 
corresponds to a negative strange-quark polarization
\begin{equation}
\Delta s_{Q^2 \rightarrow \infty}
=
\frac{1}{3}
(g_A^{(0)}|_{\rm pDIS, Q^2 \rightarrow \infty} - g_A^{(8)})
=
- 0.08 \pm 0.01 ({\rm stat.}) \pm 0.02 ({\rm syst.})
\end{equation}
-- that is,
polarized in the opposite direction to the spin of the proton.

What physics separates the values of the octet and singlet axial-charges ?

\section{Spin and the singlet axial-charge $g_A^{(0)}$}

There are two key issues:
the physics interpretation of the flavour-singlet axial-charge $g_A^{(0)}$
and
possible SU(3) breaking in the extraction of $g_A^{(8)}$ from hyperon
$\beta$-decays.

First consider $g_A^{(0)}$.
Gluonic information feeds into $g_A^{(0)}$ through the QCD axial anomaly.
QCD theoretical analysis leads to the formula~\cite{bassrmp,ar,et,ccm,bint}
\begin{equation}
g_A^{(0)}
=
\biggl(
\sum_q \Delta q - 3 \frac{\alpha_s}{2 \pi} \Delta g \biggr)_{\rm partons}
+ {\cal C}_{\infty}
.
\label{eqa10}
\end{equation}
Here $\Delta g_{\rm partons}$ is the amount of spin carried
by polarized gluons in the polarized proton
($\alpha_s \Delta g \sim {\rm constant}$ as
 $Q^2 \rightarrow \infty$~\cite{ar,et})
and
$\Delta q_{\rm partons}$ measures the spin carried by quarks
and
antiquarks
carrying ``soft'' transverse momentum $k_t^2 \sim P^2, m^2$
where
$P$ is a typical gluon virtuality
and
$m$ is the light quark mass.
The polarized gluon term is associated with events in polarized
deep inelastic scattering where the hard photon strikes a
quark or antiquark generated from photon-gluon fusion and
carrying $k_t^2 \sim Q^2$~\cite{ccm,bint}.
${\cal C}_{\infty}$ denotes a potential non-perturbative gluon
topological contribution
which is associated with
the possible subtraction
constant in the dispersion relation for $g_1$
and Bjorken $x=0$~\cite{bassrmp}:
$g_A^{(0)}|_{\rm pDIS} = g_A^{(0)} - {\cal C}_{\infty}$.

The subtraction constant, if finite, is a non-perturbative effect and
vanishes in perturbative QCD.
It is sensitive to the mechanism of axial U(1) symmetry breaking and
the realization
of axial U(1) symmetry breaking by instantons:
spontaneous U(1) symmetry breaking by instantons
naturally generates a subtraction constant whereas explicit symmetry
breaking does not \cite{topology}.
The QCD vacuum is a Bloch superposition of states
characterised
by non-vanishing topological winding number and non-trivial chiral
properties.
When we put a valence quark into this vacuum it can act as a
source which polarizes the QCD 
vacuum with net result that the spin ``dissolves'' and some
fraction of
the spin of the constituent quark
is associated with non-local gluon topology with support only at
Bjorken $x=0$.

Possible explanations for the small value of $g_A^{(0)}|_{\rm pDIS}$
extracted
from the polarized deep inelastic experiments
include
screening from positive gluon polarization,
negative strangeness polarization in the nucleon,
a subtraction at infinity in the dispersion relation for $g_1$
associated with non-perturbative gluon topology
and
connections to axial U(1) dynamics \cite{tgv,shore,hf,bass99},
as well as possible SU(3) breaking in $g_A^{(8)}$
-- possibly as large as 20\%\cite{sb2010,Jaffe:1990a}.
The QCD axial anomaly decouples from the non-singlets $g_A^{(3)}$ and
$g_A^{(8)}$.

One would like to understand the dynamics which appears to suppress
the singlet axial-charge extracted from polarized deep inelastic
scattering relative to the OZI prediction $g_A^{(0)} = g_A^{(8)}$
and also the sum-rule for the longitudinal spin structure
of the nucleon
\begin{equation}
\frac{1}{2} = \frac{1}{2} \sum_q \Delta q + \Delta g + L_q + L_g
\end{equation}
where $L_q$ and $L_g$ denote the orbital angular momentum contributions.
There is presently a vigorous programme to disentangle the different
contributions. 
Key experiments include semi-inclusive 
polarized deep inelastic scattering (COMPASS and HERMES) 
and polarized proton-proton collisions (PHENIX and STAR at RHIC),
as
well as deeply virtual Compton scattering to learn about total
angular momentum.

\section{The shape of $g_1$}

To understand the proton spin puzzle, it is interesting to look at
the $x$ dependence of the measured $g_1$ spin structure function.
Deep inelastic measurements of $g_1$ have been performed in experiments at
CERN, DESY, JLab and SLAC.
There is a general consistency among all data sets.
COMPASS are yielding precise new data at small $x$, down to $x \sim 0.004$.
JLab are focussed on the large $x$ region.

Precise measurements of the deuteron spin structure 
function $g_1^d$
show the remarkable feature that $g_1^d$
is consistent with zero in the small $x$ region between 0.004 and 0.02
\cite{compassnlo}.
In contrast, 
the isovector part of $g_1$ is observed to rise at small $x$ as 
$\sim x^{-0.22 \pm 0.07}$
and is much bigger
than the isoscalar part of $g_1$\cite{compass10}.
This is in
sharp contrast to the situation in the unpolarized structure
function $F_2$ where the small $x$ region is dominated by isoscalar
pomeron exchange. 
The $g_1^{p-n}$ data are consistent with
quark model and perturbative QCD predictions in the valence region
$x > 0.2$ \cite{epja}. 
The size of $g_A^{(3)}$ 
forces us to accept a
large contribution from small $x$ and 
the observed rise in $g_1^{p -n}$
is in excellent agreement with the prediction 
$g_1^{p-n} \sim  x^{-0.22}$
of
hard Regge exchange
- in particular a possible $a_1$ hard-pomeron cut\cite{bassmb}
involving the hard-pomeron which
seems 
to play an important role in unpolarized deep inelastic scattering\cite{pvl}.

The ``missing spin'' is associated with a ``collapse''
in the isosinglet part of $g_1$ to something close to
zero instead of a valence-like rise
for $x$ less than about 0.02.
This isosinglet part is the sum of
SU(3)-flavour singlet and octet contributions.
If there were a large positive polarized gluon contribution
to the proton's spin, this would act to
drive the small $x$ part of the singlet part of $g_1$ negative\cite{bt91}
--
that is, acting in the opposite direction to any valence-like
rise at small $x$.
However, gluon polarization measurements at COMPASS, HERMES and RHIC
constrain this spin contribution to be small in measured kinematics
meaning that the sum of valence and sea quark contributions
is suppressed at small $x$.
(Soft Regge theory predicts that the singlet term should
 behave as
 $\sim N \ln x$ in the small $x$ limit,
 with the coefficient $N$ to be determined from experiment\cite{sbpvl,fec94}.)

There is presently a vigorous programme to disentangle the different
contributions involving experiments in semi-inclusive polarized deep
inelastic scattering and polarized proton-proton
collisions~\cite{sb2009,Mallot:2006,hermess,compasss}.
These direct measurements show no evidence for negative polarized
strangeness in
the region $x> 0.006$ 
(in apparent contrast to the extraction of negative strangeness
 polarization extracted from inclusive measurements of $g_1$).
For gluon polarization,
present measurements
suggest
$|- 3 \frac{\alpha_s}{2 \pi} \Delta g| < 0.06$
corresponding to $|\Delta g| < 0.4$ with $\alpha_s \sim 0.3$.
That is,
they are not able to account
for the difference
$( g_A^{(0)}|_{\rm pDIS} - g_A^{(8)} ) \sim -0.25$
obtained via Eq.(4).
An independent measurement of the strange-quark axial-charge
could be made through neutrino-proton elastic scattering\cite{bcsta}.
The axial-charge measured in $\nu p$ elastic scattering 
is independent of
any assumptions about the presence or absence of a subtraction at
infinity in the dispersion relation for $g_1$ and the $x \sim 0$
behaviour of $g_1$.
Further measurements to push the small $x$ frontier in polarized 
deep inelastic scattering would be possible with a polarized $ep$ 
collider\cite{trento}.

\section{SU(3) breaking and $g_A^{(8)}$}

\begin{table}[b!]
\tbl{$g_A/g_V$ from $\beta$-decays with $F=0.46$ and $D=0.80$,
together with the mathematical form predicted
in the MIT Bag with
effective colour-hyperfine interaction (see text and Ref.[32]).}
{\begin{tabular}{llllr}
Process             &  measurement      &  SU(3) combination & Fit value
& MIT + OGE
\\
\hline
$n \rightarrow p$     &  $1.270 \pm 0.003$  &  $F+D$   & 1.26
& $\frac{5}{3} B' + G$
\\
$\Lambda^0 \rightarrow p$ & $0.718 \pm 0.015$ & $F+\frac{1}{3}D$ & 0.73
& $B'$
\\
$\Sigma^- \rightarrow n$  & $-0.340 \pm 0.017$ & $F-D$ & -0.34
& $-\frac{1}{3} B' - 2G$
\\
$\Xi^- \rightarrow \Lambda^0$ & $0.25 \pm 0.05$ & $F-\frac{1}{3}D$ & 0.19
& $\frac{1}{3} B' - G$
\\
$\Xi^0 \rightarrow \Sigma^+$ & $1.21 \pm 0.05$ & $F+D$ & 1.26
& $\frac{5}{3} B' + G$
\\
\hline
\end{tabular}}
\end{table}

Given that the contributions to $g_A^{(0)}$ from the measured distribution
$\Delta s$ and from $-3\frac{\alpha_s}{2 \pi} \Delta g$
are small,
it is worthwhile to ask about the value of $g_A^{(8)}$.
The value $0.58$
is extracted from a 2 parameter fit to hyperon $\beta$-decays
in terms of the SU(3) constants $F=0.46$ and $D=0.80$~\cite{fec}
-- see Table 1.
The fit is good to $\sim 20\%$ accuracy~\cite{Jaffe:1990a,leaders}.
The uncertainty quoted
for $g_A^{(8)}$ has been a matter of some debate. There is
considerable evidence that SU(3) symmetry may be badly broken
and some have suggested that the error on $g_A^{(8)}$ should be
as large as 25\%~\cite{Jaffe:1990a}.
More sophisticated fits will also include chiral corrections.
Calculations of non-singlet axial-charges in relativistic
constituent quark models
are sensitive
to the confinement potential,
effective colour-hyperfine interaction~\cite{myhrer,myhrer2,awt,Close:1978},
pion and kaon clouds
plus additional wavefunction corrections~\cite{schreiber}
chosen to reproduce the physical value of $g_A^{(3)}$.

This physics has recently been investigated
by Bass and Thomas~\cite{sb2010}
within the Cloudy Bag model (CBM)~\cite{schreiber,Kazuo}
which has the attractive feature
that when pion cloud and quark mass effects are turned off the model
reproduces the SU(3) analysis.
One finds that chiral corrections significantly reduce the value of
$g_A^{(8)}$. This, in turn, has the effect of increasing 
the value of $g_A^{(0)}|_{\rm pDIS}$ and consequently reducing 
the absolute value of
the ``polarized strangeness''
extracted from inclusive polarized deep inelastic scattering.

The Cloudy Bag~\cite{Thomas:1984}
was designed to model confinement and spontaneous chiral
symmetry breaking, taking into account pion physics and the manifest
breakdown of chiral symmetry at the bag surface in the MIT bag.
If we wish to describe proton spin data including matrix elements of
$J_{\mu 5}^3$, $J_{\mu 5}^8$ and $J_{\mu 5}$,
then we would like to know that the model versions of these currents
satisfy the relevant Ward identities.
For the scale-invariant 
non-singlet 
axial-charges $g_A^{(3)}$ and $g_A^{(8)}$,
corresponding to the matrix elements of partially conserved
currents, the model is well designed to make a solid prediction.

The effective colour-hyperfine interaction has the quantum numbers of
one-gluon exchange (OGE).
In models of hadron spectroscopy this interaction
plays an important role in the nucleon-$\Delta$ and $\Sigma-\Lambda$
mass differences,
as well as the nucleon magnetic moments~\cite{Close:1978} and the spin
and flavor dependence of parton distribution functions~\cite{Close:1988br}.
It shifts total angular-momentum between spin and orbital
contributions and, therefore, also contributes to model calculations of
the octet axial-charges~\cite{myhrer,myhrer2,awt}.
In Bag model calculations 
one also needs to include 
wavefunction corrections
associated with the well known issue that,
for the MIT and Cloudy Bag models,
the nucleon wavefunction is not translationally invariant and 
the
centre of mass is not fixed.
To compare the model results with experiment
we take the view~\cite{schreiber}
that, in principle,
the model - with corrections -
should give the experimental value of $g_A^{(3)}$.
We therefore choose the centre-of-mass factor 
phenomenologically
to give the experimental value of $g_A^{(3)}$.
This then fixes the parameters of the model and
allows us to use it to make a model prediction for $g_A^{(8)}$.

Without pion cloud corrections the MIT Bag with centre of mass corrections
reproduces the SU(3) analysis of the axial-charges extracted 
from $\beta$-decays.
This is illustrated in Table 1.
Without additional physics input,
e.g. pion chiral corrections,
there is a simple algebraic relation between the SU(3) 
parameters $F$ and $D$,
the bag parameter $B^\prime$ and the OGE correction $G$:
$F = \frac{2}{3} B' - \frac{1}{2} G$ 
and
$D = B' + \frac{3}{2} G$.
The numerical agreement is very good~\cite{sb2010}.

The pion cloud of the nucleon also renormalizes the nucleon's 
axial-charges by shifting intrinsic spin into orbital
angular momentum~\cite{myhrer,schreiber}.
In the Cloudy Bag Model (CBM)~\cite{Thomas:1984},
the nucleon wavefunction is written as a Fock expansion
in terms of a bare MIT nucleon,
$|{\rm N}\rangle$, and baryon-pion, $|{\rm N} \pi \rangle$ and
$|{\Delta \pi}\rangle$, Fock states.
The probabilities to find the nucleon in each Fock component
are determined phenomenologically 
by fitting to a wealth of nucleon observables~\cite{awtpion}.
The expansion converges rapidly and we may safely truncate the Fock
expansion at the one pion level.
When we calculate the pion and kaon cloud chiral corrections 
to $g_A^{(8)}$
we also have to 
choose the chiral representation, in particular whether to use the 
original surface coupling or the later volume coupling
version of the Cloudy Bag model.

The extent of the reduction in $g_A^{(8)}$
depends upon the
version of the CBM used, lying in the
range $0.49 \pm 0.02$ for the original CBM and
$0.42 \pm 0.02$ for the volume coupling version~\cite{sb2010}.
These changes alone raise the value of
$g_A^{(0)}|_{\rm pDIS, Q^2 \rightarrow \infty}$
derived from the experimental data
from
$0.33 \pm 0.03 ({\rm stat.}) \pm 0.05 ({\rm syst.})$ to
$0.35 \pm 0.03 ({\rm stat.}) \pm 0.05 ({\rm syst.})$ and
$0.37 \pm 0.03 ({\rm stat.}) \pm 0.05 ({\rm syst.})$, respectively.
Both of these values
have the effect of reducing the level of OZI violation
associated with the difference
$g_A^{(0)}|_{\rm pDIS} - g_A^{(8)}$
from $-0.25 \pm  0.07$
to
just $-0.14 \pm 0.06$
and $-0.05 \pm 0.06$, respectively.
It is this OZI violation which eventually needs to be explained
in terms of singlet degrees of freedom:
effects associated with polarized glue and/or a topological effect
associated with $x=0$.

The uncertainty in this model calculation
lies in the small
ambiguity between the two chiral representations that one can choose.
In order to quote an overall value that properly encompasses these
possibilities
we follow the Particle Data Group procedure~\cite{PDG_proc}
for combining data that {\it may} not be compatible to estimate the overall
error,
finding a combined value of $g_A^{(8)} = 0.46 \pm 0.05$
(with the corresponding semi-classical
 singlet axial-charge or spin fraction
being $0.42 \pm 0.07$ before inclusion of gluonic effects)~\cite{sb2010}.
Note that
the error $\pm 0.02$ on $g_A^{(8)}$
quoted for each version of the model follows from varying
over the phenomenological range of possible pion parameters within first order
perturbation theory. In terms of analogy to experimental errors, the $\pm 0.02$
is like a statistical error and the final $\pm 0.05$ error includes
systematic effects.
With this final value for
$g_A^{(8)}$ the corresponding experimental value of 
$g_A^{(0)}|_{\rm pDIS}$
would increase to $g_A^{(0)}|_{\rm pDIS} = 0.36 \pm 0.03 \pm 0.05$.

\section{Towards possible understanding }

Where are we in our understanding of
the spin structure of the proton
and the small value of $g_A^{(0)}|_{\rm pDIS}$ ?
Measurements of valence, gluon and sea polarization suggest that
the polarized glue term
$-3 \frac{\alpha_s}{2 \pi} \Delta g_{\rm partons}$
and strange quark contribution
$\Delta s_{\rm partons}$
in Eq.(6)
are unable to resolve the small value of $g_A^{(0)}|_{\rm pDIS}$.
Two explanations are suggested 
within the theoretical and experimental uncertainties depending
upon the magnitude of SU(3) breaking in the nucleon and hyperon
axial-charges. 
One is a value of $g_A^{(8)} \sim 0.5$ 
(as suggested by the surface coupling model) 
plus an axial U(1)
topological effect at $x=0$
associated with a finite subtraction constant in the $g_1$ 
dispersion relation. 
The second is a much larger pion cloud reduction of $g_A^{(8)}$
to a value $\sim 0.4$ 
(as suggested by the volume coupling model in first order pion
 cloud perturbation theory).
Combining the theoretical error on the pion cloud chiral corrections
embraces both possibilities.
The proton spin puzzle seems to be telling us about
the interplay of valence quarks 
with chiral dynamics and the complex vacuum structure of QCD.

\section*{Acknowledgments}
The research of
SDB is supported by the Austrian Science Fund (FWF grant P20436).
I thank A. W. Thomas for collaboration on physics reported
here and  
H. Fritzsch for the invitation to this stimulating meeting
in honour of the 80th birthday of Prof. Murray Gell-Mann.

\bibliography{sample}

\end{document}